\documentclass{appolb}
\usepackage{graphicx}
\usepackage{cite}
\usepackage[tight]{subfigure}    

\begin{document}
\title{PHENIX results on L\'evy analysis of Bose-Einstein correlation functions%
\thanks{Presented at CPOD 2016: Critical Point and Onset of Deconfinement, Wroc\l{}aw, Poland, May 30 - June 4, 2016
}
}
\author{D\'aniel Kincses for the PHENIX Collaboration
\address{E\"otv\"os Lor\'and University, P\'azm\'any P\'eter s\'et\'any 1/A, Budapest, 1117 Hungary}
}
\pagestyle{plain}
\maketitle
\begin{abstract}

The nature of the quark-hadron phase transition can be  investigated through analyzing the space-time structure of the hadron emission  source. For this, the Bose-Einstein or HBT correlations of identified charged particles are among the best observables. In this paper we present the latest results from the RHIC PHENIX experiment on such measurements.

\end{abstract}
\PACS{25.75.Dw}

\section{Introduction}

The PHENIX experiment at the BNL Relativistic Heavy Ion Collider (RHIC) has collected comprehensive data in multiple different collision systems from p+p, p+A through A+A up to U+U collisions, at energies that are varied in the region where the change from first order to crossover phase transition is expected to occur. The importance of the RHIC beam energy scan program lies in the possibility of  investigating the phase diagram of QCD matter, and the quark-hadron phase transition. One of the best tools to gain information about the particle-emitting source is the measurement of Bose-Einstein or HBT correlations of identical bosons. In our  latest measurements, we utilize L\'evy-type sources~\cite{Csorgo:2003uv,Csorgo:2009gb} to describe the measured  correlation functions. In case of a second order QCD phase-transition, one of the source parameters, the index of stability  $\alpha$ is related to one of the critical exponents (the so-called correlation  exponent $\eta$). Thus Bose-Einstein correlation data may yield information on the nature of the quark-hadron  phase transition, particularly it may shed light on the location of the critical  endpoint (CEP) on the phase-diagram. 

\newpage
\section{Beam energy dependence of HBT radii}

\begin{figure}[htb]
\centerline{%
\includegraphics[width=8cm]{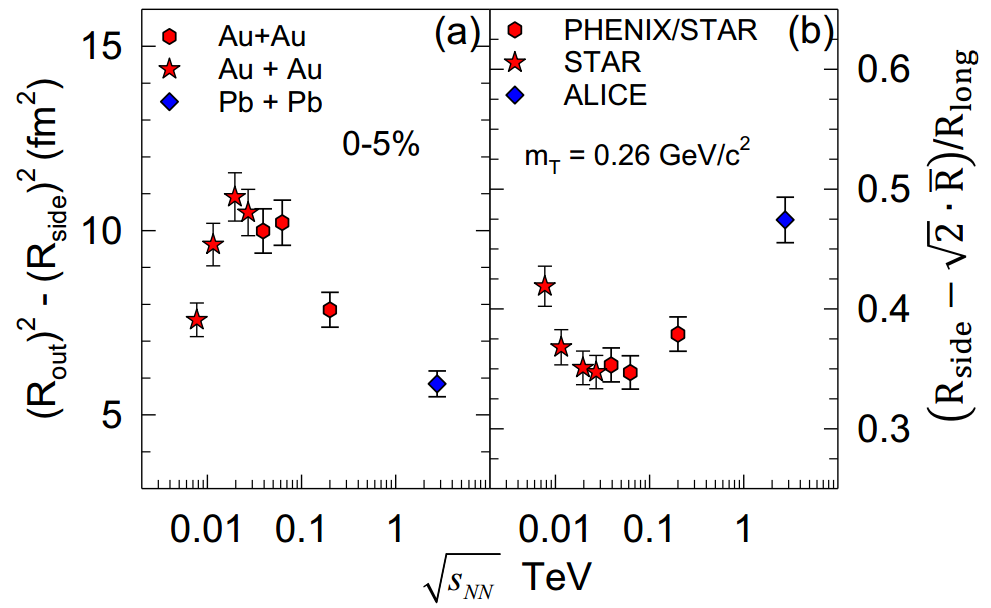}}
\caption{The $\sqrt{s_{NN}}$ dependencies of combinations of HBT radii sensitive to expansion velocity  (a) and emission duration (b) from \cite{Adare:2014qvs,Adamczyk:2014mxp,Kisiel:2011}.}
\label{Fig1}
\end{figure}

Today, high energy physics experiments measure the scale parameter of HBT correlation functions (often called HBT radii) as a function of particle type, transverse momentum, azimuthal angle, collision energy, and collision geometry. Recently, PHENIX measured the Gaussian HBT radii of two-pion Bose-Einstein correlation functions in Au+Au collisions at several beam energies \cite{Adare:2014qvs}. The extracted radii, which were compared to recent STAR \cite{Adamczyk:2014mxp} and ALICE \cite{Kisiel:2011} data, show characteristic scaling patterns as a function of the transverse mass of the emitted pion pairs, consistent with hydrodynamic like expansion. On Fig. \ref{Fig1}. we show specific combinations of the three-dimensional radii \cite{Adare:2014qvs,Adamczyk:2014mxp,Kisiel:2011} that are sensitive to the medium expansion velocity and lifetime, and the pion emission duration. These show non-monotonic $\sqrt{s_{NN}}$ dependencies, which may be an indication of the critical endpoint in the phase diagram of hot and dense nuclear matter. 

\section{A possible way of finding the critical point}

If we leave behind the Gaussian assumption for the correlation shape, there may be an other way to search for the place of the critical point on the phase diagram, connected to the non-Gaussian shape of the correlation functions. The generalization of the Gaussian is the so-called L\'evy-distribution \cite{Csorgo:2003uv,Csorgo:2009gb}:
\begin{equation}
\displaystyle \mathcal{L}(\alpha,R,r)=\frac{1}{(2\pi)^3} \int d^3q e^{iqr} e^{-\frac{1}{2}|qR|^{\alpha}},
\end{equation}
where the $\alpha = 2$ case corresponds to the Gaussian, and the $\alpha=1$ to the Cauchy distribution. With this assumption for the shape of the source, the shape of the correlation functions then becomes
\begin{equation}
\large{C_{2}(k)=1+\lambda\cdot e^{-(2Rk)^{\alpha}}},
\label{C0}
\end{equation}
where $k$ is the absolute value of half of the relative momentum. From statistical physics we know that the definition of critical exponent $\eta$ comes from the spatial correlations being proportional to $r^{-(d-2+\eta)}$, where $d$ represents the number of dimensions. In case of symmetrical L\'evy-stable distributions, the spatial correlations are proportional to $r^{-1-\alpha}$. So it turns out, that L\'evy exponent $\alpha$ is identical to critical exponent $\eta$. The value of $\eta$ at the CEP can be predicted if we assume that the QCD universality class is the same as the universality class of the 3 dimensional Ising model \cite{Halasz:1998qr, Stephanov:1998dy}: $\eta = 0.03631(3)$ \cite{El-Showk:2014dwa}. In case of random field 3D Ising model, this value is somewhat different: $\eta = 0.50\pm0.05$ \cite{Rieger:1995aa}. If we measure the $\alpha$ parameter at different beam energies, we may gain insight into the phase diagram of QCD matter, as the change in its behaviour is connected to the proximity of the CEP.

\section{L\'evy analysis of HBT correlation functions}

\begin{figure}[h!]
\centerline{%
\includegraphics[width=0.9\textwidth]{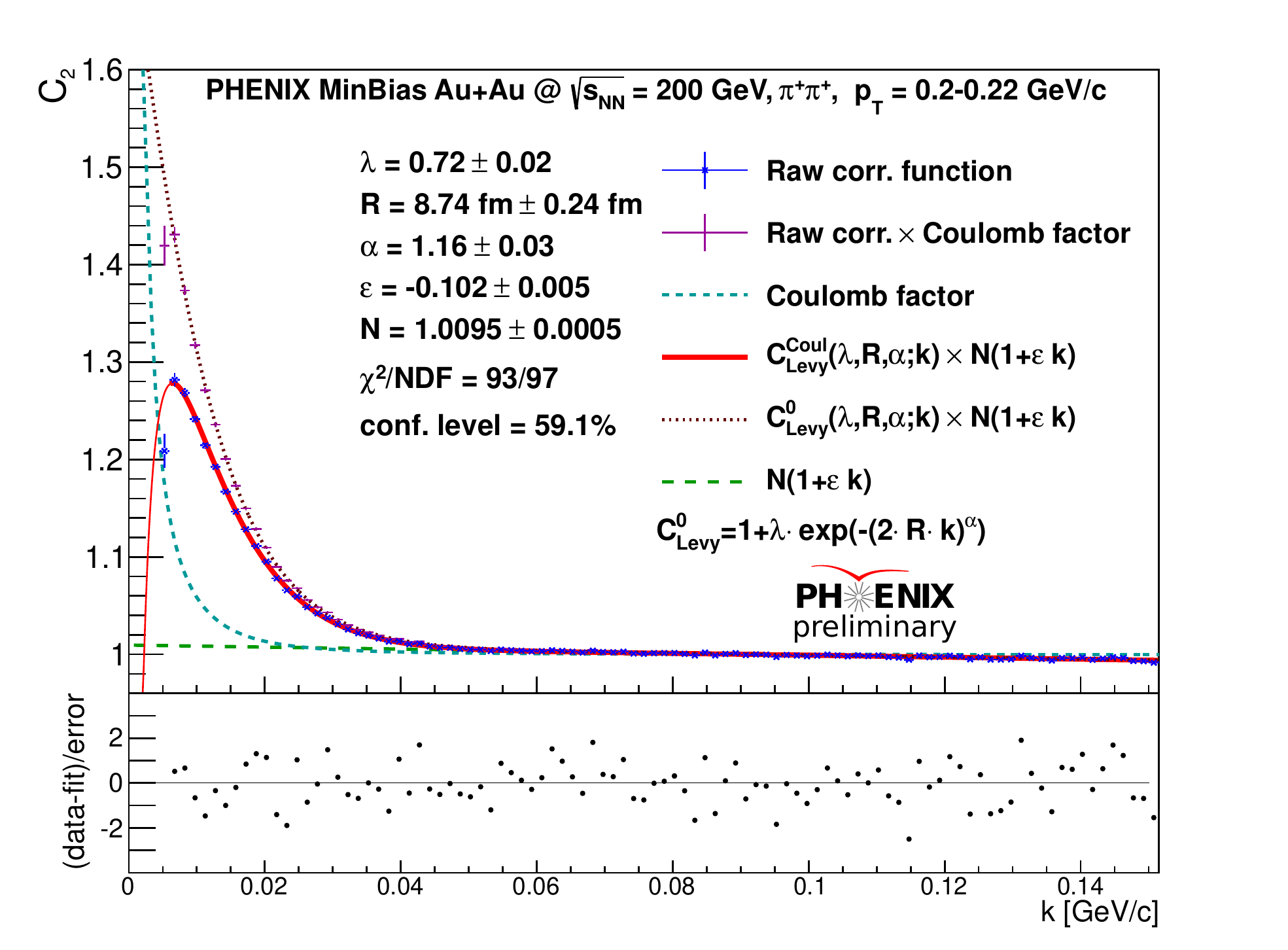}}
\caption{Example fit of a Bose-Einstein correlation function of $\pi^+\pi^+$ pairs with average $p_T$ between 0.2 and 0.22 GeV/c measured in the longitudinal co-moving frame. The fit shows the measured correlation function and the complete fit function, while a ``Coulomb-corrected" fit function $C^{(0)}(k)$ is also shown, with the data multiplied by $C^{(0)}/C^{Coul}$. In this analysis we measured 62 such correlation functions (for ++ and - - pairs, in 31 $p_T$ bins).} 
\label{Fig2}
\end{figure}

The data sample used in the latest analysis consists of $\sim$7 billion minimum bias Au+Au events at $\sqrt{s_{NN}} = 200$ GeV recorded by PHENIX during Run-10. We measured two-pion correlation functions for $\pi^+\pi^+$ and $\pi^-\pi^-$ pairs in 31 $p_T$ bins (ranging from 180 MeV/c to 850 MeV/c), where $p_T$ is the average transverse momentum of the pair. If one uses a spherically symmetric L\'evy type source and couples it with the core-halo model, then the shape of the two-particle correlation function takes the form seen in Eq. (\ref{C0}). However, fitting this functional form does not yield meaningful physics results since it does not take into account the final state Coulomb interaction. To handle this, we have incorporated the effect of the Coulomb repulsion of the identically charged pion pairs into the fit function. An example fit is shown on Fig. \ref{Fig2}. 

\begin{figure}%
\vspace{-20pt}
\centerline{
\newskip\subfigcapskip \subfigcapskip = -45pt
\subfigure[][]{%
\label{fig:3-a}%
\includegraphics[width=0.62\textwidth]{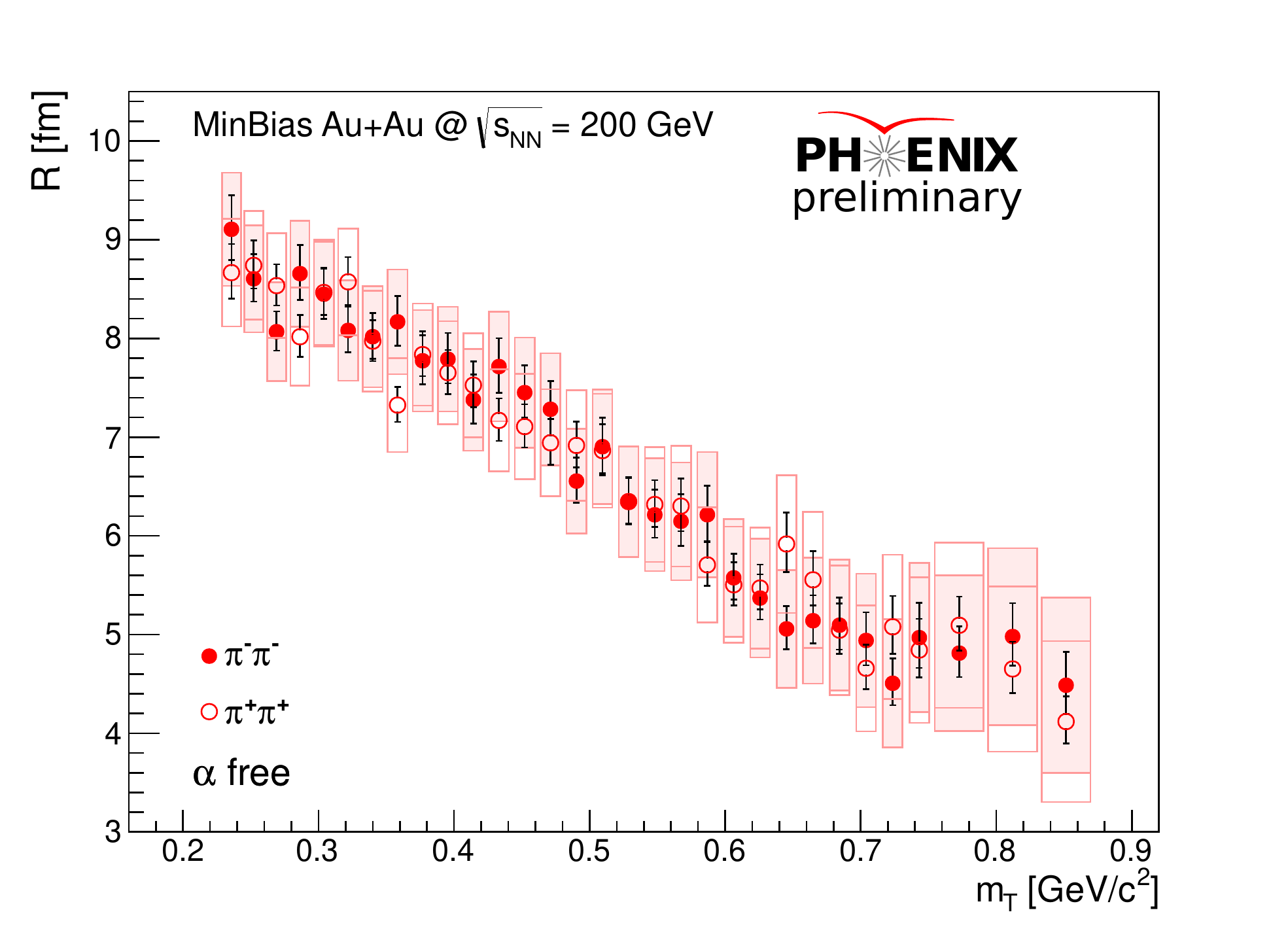}}%
\hspace{-20pt}%
\subfigure[][]{%
\label{fig:3-b}%
\includegraphics[width=0.62\textwidth]{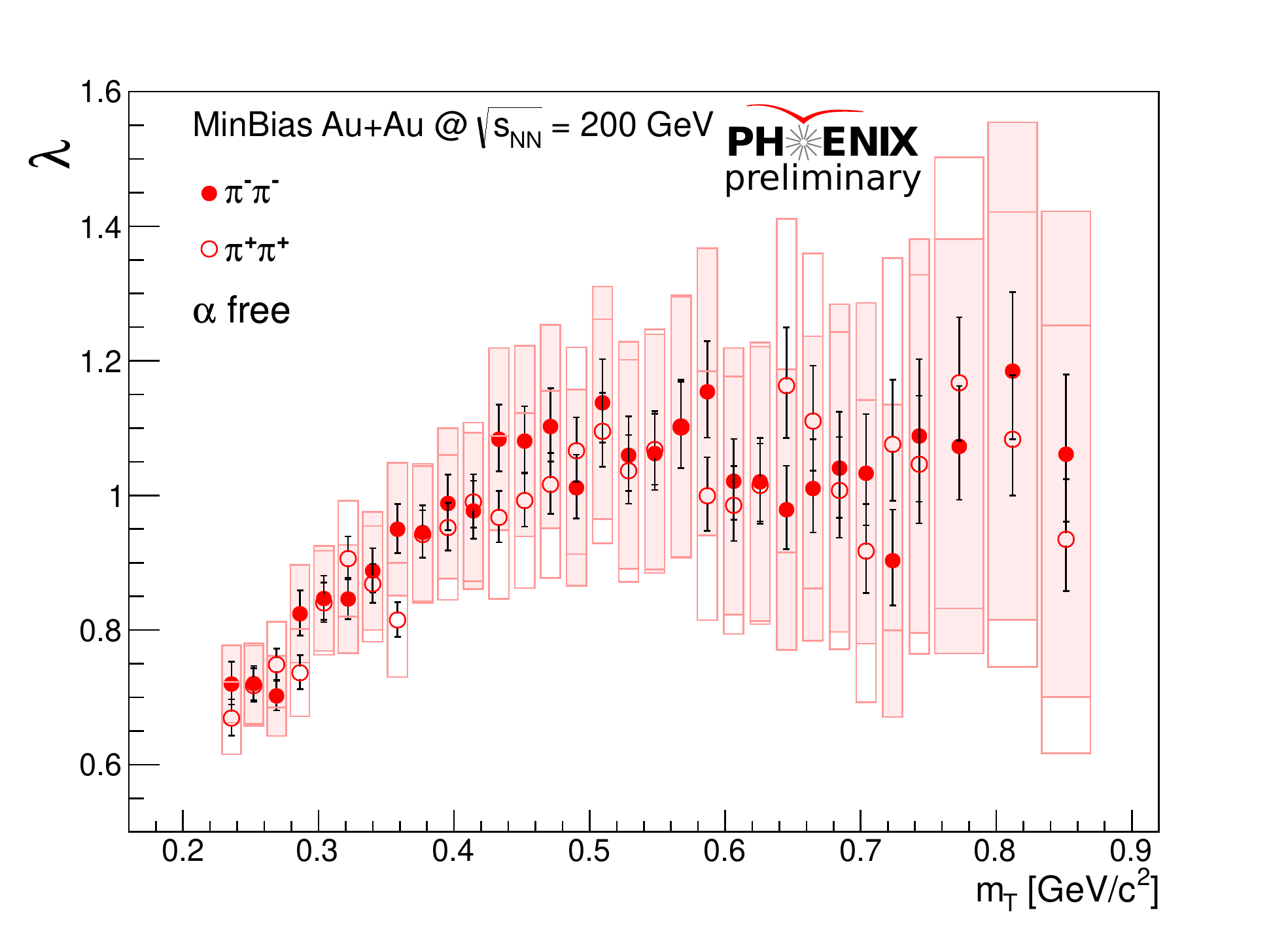}}\\}
\newskip\subfigcapskip \subfigcapskip = -130pt
\vspace{-3pt}
\centerline{
\subfigure[][]{%
\label{fig:3-c}%
\includegraphics[clip, trim=0cm 0cm 0cm 1.25cm,width=0.62\textwidth]{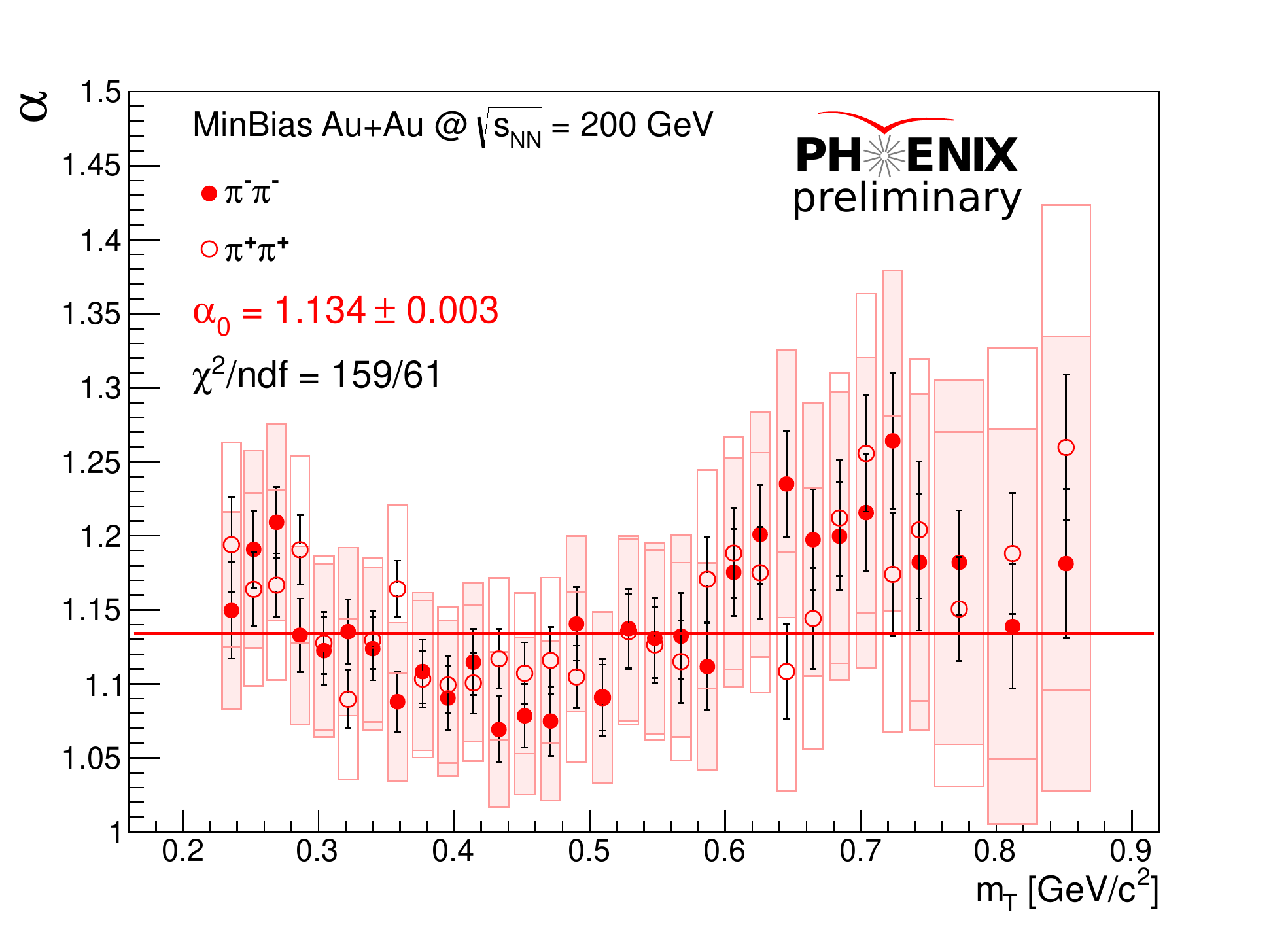}}%
\hspace{-20pt}%
\subfigure[][]{%
\label{fig:3-d}%
\includegraphics[clip, trim=0cm 0cm 0cm 1.25cm,width=0.62\textwidth]{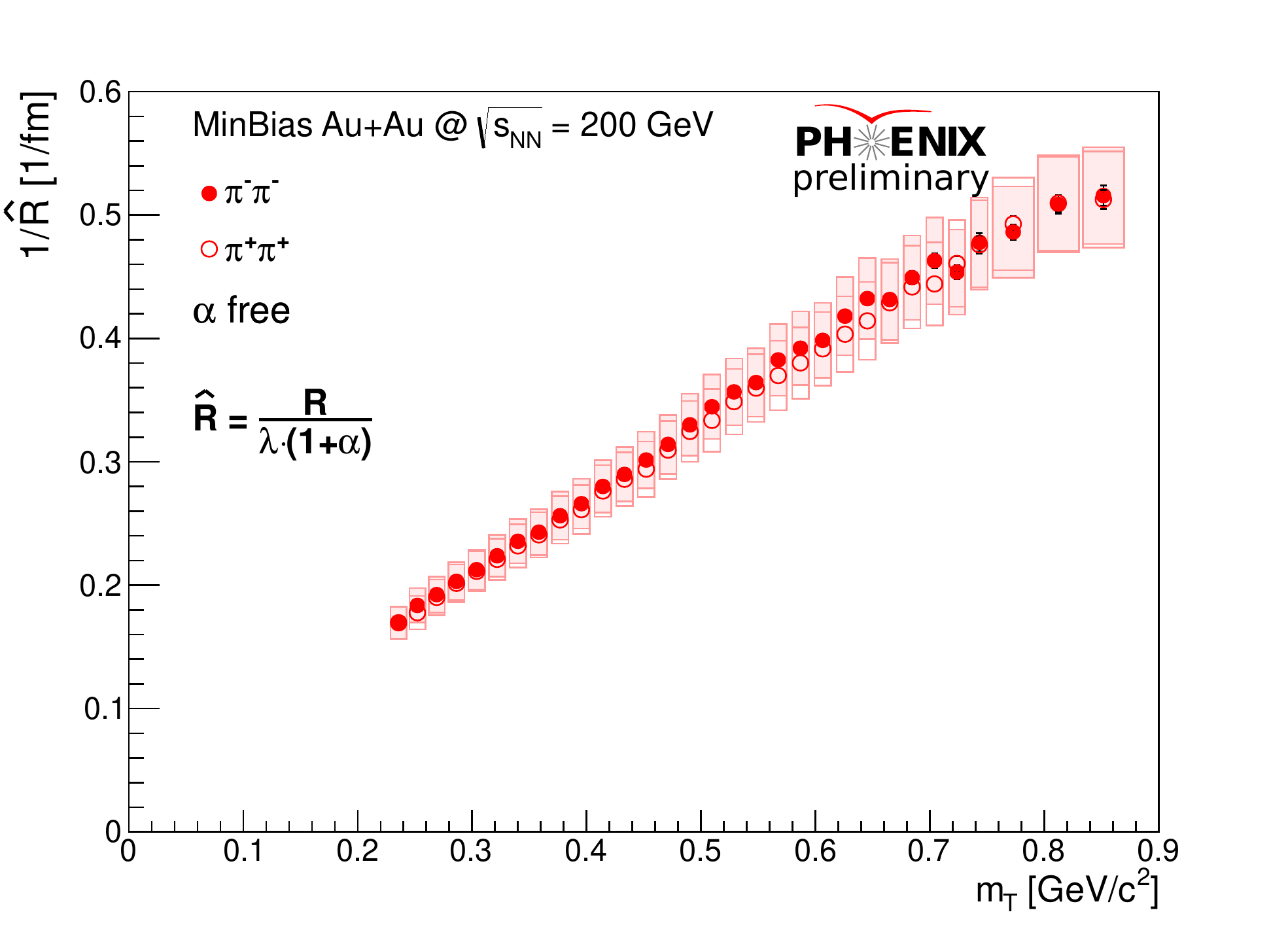}}}%
\caption[]{Fit parameters versus average $m_T$ of the pair with statistical and symmetric systematic uncertainties shown as bars and boxes, respectively. 
\subref{fig:3-a} L\'evy scale parameter R
\subref{fig:3-b} Correlation strength parameter $\lambda$
\subref{fig:3-c} L\'evy exponent $\alpha$
\subref{fig:3-d} Newly found scale parameter $1/\widehat{R}$}%
\label{Fig3}%
\vspace{-15pt}
\end{figure}

On Fig. \ref{Fig3}. we  present the resulting physical fit parameters ($\lambda, R, \alpha$), versus pair $m_T$ (corresponding to the given $p_T$ bin), as well as a newly found scale parameter, $1/\widehat{R}=(\lambda(1+\alpha))/R$. On Fig. \ref{Fig3} \subref{fig:3-a} we can see that the $m_T$ dependence of R shows the usual decreasing trend predicted by hydrodynamics for an expanding source. Fig. \ref{Fig3} \subref{fig:3-b} shows that the strength parameter $\lambda$ seems to saturate at high $m_T$, and a decreasing trend is clearly visible with decreasing $m_T$. The value of $\alpha$ (shown in Fig. \ref{Fig3} \subref{fig:3-c})  is approximately independent of $m_T$ and it is far from the Gaussian limit ($\alpha=2$) and the 3D Ising limit ($\alpha\leq0.5$) corresponding to the conjectured value at the CEP. Regarding $U_A(1)$ symmetry restoration~\cite{Csorgo:2009pa}, our new preliminary results are consistent with earlier preliminary results~\cite{Csanad:2005nr}  within  statistical uncertainties. The three physical fit parameters are strongly correlated (the value of the correlation coefficients between  ($\lambda,R$), ($R,\alpha$) are above 90$\%$), so we searched for parameter combinations that represent the fit function in a more unambiguous way. We indeed found such a parameter: $\widehat{R}=R/(\lambda(1+\alpha))$. If we calculate $R$ as $R=\widehat{R}\lambda(1+\alpha)$ and use this new parameter for the fitting, the correlation coefficients between ($\lambda,\widehat{R}$), ($\widehat{R},\alpha$) drop to  $20$-$30\%$. Fig. \ref{Fig3} \subref{fig:3-d} indicates that $1/\widehat{R}$ is approximately linear in $m_T$. The modified L\'evy scale parameter $\widehat{R}$ was found experimentally, but, as of now, the physical interpretation remains an open question. However, now that the method of this measurement is well established, the next step is the analysis of the lower energy data sets - this may lead us forward in exploring  the phase diagram of hot and dense, strongly interactive matter.
\vspace{10pt}
  
  This research was supported by the funding agencies listed in ref. \cite{funding}. In addition, D. K. was supported through the New National Excellence program of the Hungarian Ministry of Human Capacities.
\newpage

\end{document}